\documentclass[aps,prl,preprint,superscriptaddress]{revtex4-2}

\usepackage{graphicx}
\usepackage{dcolumn}
\usepackage{bm}
\usepackage{color}
\usepackage{float}

\begin{document}

\title{Direct evidence for efficient carrier multiplication in the topological insulator Bi$_2$Se$_3$}

\author{Michael Herb}
\affiliation{Department for Experimental and Applied Physics, University of Regensburg, Regensburg, Germany}
 
\author{Leonard Weigl}
\affiliation{Department for Experimental and Applied Physics, University of Regensburg, Regensburg, Germany}

\author{Niklas Hofmann}
\affiliation{Department for Experimental and Applied Physics, University of Regensburg, Regensburg, Germany}
 
\author{Johannes Gradl}
\affiliation{Department for Experimental and Applied Physics, University of Regensburg, Regensburg, Germany}

\author{Jason F. Khoury}
\affiliation{Princeton University, USA}
\affiliation{School of Molecular Sciences, Arizona State University, Tempe, AZ, 85282, USA}

\author{Leslie Schoop}
\affiliation{Princeton University, USA}

\author{Isabella Gierz}
\email[]{isabella.gierz@ur.de}
\affiliation{Department for Experimental and Applied Physics, University of Regensburg, Regensburg, Germany}

\date{\today}

\begin{abstract}

Carrier multiplication (CM), where the absorption of a single photon results in the generation of several electron-hole pairs via impact ionization, plays a pivotal role in the quest for enhancing the performance of solar cells beyond the Shockley-Queisser limit. The combination of its narrow bandgap relative to the photon energy of visible light, along with its low phonon frequencies that hinder efficient energy dissipation into phonons, makes the topological insulator Bi$_2$Se$_3$ an optimal candidate material for efficient CM. Here we use time- and angle-resolved photoemission spectroscopy (trARPES) to trace the number of electron-hole pairs after photoexcitation of Bi$_2$Se$_3$ with visible pump pulses at $\hbar\omega=2$\,eV. We find that both the number of electrons inside the conduction band as well as the number of holes inside the valence band keep increasing long after the pump pulse is gone, providing direct evidence for CM. We also analyze the transient band structure as well as the hot carrier dynamics inside the conduction band, providing a complete picture of the non-equilibrium carrier dynamics in photoexcited Bi$_2$Se$_3$ which can now serve as a basis for novel optoelectronic applications.

\end{abstract}

\maketitle



Impact ionization, where charge carriers with high kinetic energy generate electron-hole pairs upon collision with atoms, is a process that commonly occurs in semiconductors at high electric fields \cite{McKay1953, McKay1954, Wolff1954, Miller1955} or after the absorption of high-energy photons \cite{Koc1957, Tauc1959, Kolodinski1993, Brida2013, Ploetzing2014, Gierz2015}. Impact ionization plays a crucial role for the development of next-generation photodetectors and solar cells as the absorption of a single photon produces multiple electron-hole pairs which enhances the efficiency of these devices considerably. For carrier multiplication (CM) to occur, two requirements must be met: (1) The energy of the absorbed photon $\hbar\omega$ must be at least twice as big as the band gap $E_{\text{gap}}$ and (2) the excess energy $E_{\text{exc}}$ of the photoexcited electrons should not be dissipated into other degrees of freedom such as phonons. This makes the topological insulator Bi$_2$Se$_3$ with its band gap of $E_{gap}\sim300$\,meV \cite{Zhang2009, Jurczyszyn2020} (roughly an order of magnitude smaller than the photon energy of visible light) and phonon frequencies below 25\,meV \cite{Sobota2012, Sobota2014} an ideal candidate material. The non-equilibrium carrier dynamics of photoexcited Bi$_2$Se$_3$ has been excessively studied in the past. Most investigations, however, focused on the topological surface state \cite{Sobota2012, Sobota2014, Michiardi2022} and surface photovoltage (SPV) effects \cite{Hajlaoui2014, Ciocys2020, Ciocys2020, Neupane2015, Sanchez2017, Michiardi2022}. The population dynamics of the bulk bands that may provide evidence of CM in the material remain unexplored.
	
\begin{figure}
\centering
\includegraphics[width=1\columnwidth]{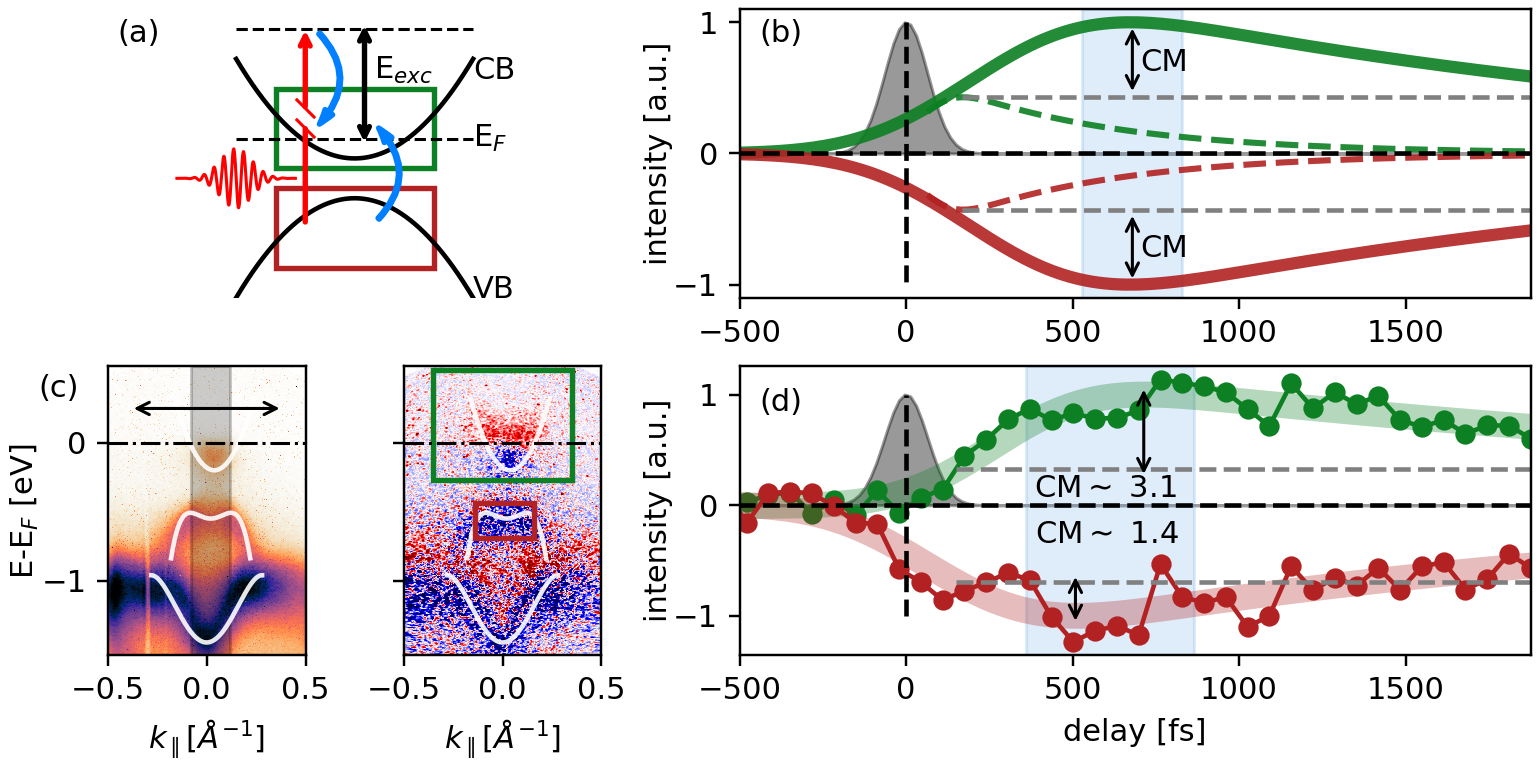}
\caption{\label{figure1} Carrier multiplication in Bi$_2$Se$_3$. (a) Schematic band structure of an n-doped semiconductor. The red arrow indicates the direct electronic transition from occupied states in the VB into unoccupied states above the Fermi level in the CB induced by photoexcitation. The blue arrows indicate carrier-carrier scattering processes where the excess energy of a photoexcited electron $E_{exc}$ in the CB is used to promote a second electron from the VB into the CB. (b) Sketch of the expected CM signature in trARPES data. Green and red curves represent the expected number of electrons and holes in the CB and VB, respectively. Solid and dashed lines indicate the expected dynamics with and without CM, respectively. (c) trARPES snapshots of Bi$_2$Se$_3$ bandstructure around the $\Gamma$-point taken at negative pump-probe delay (left) together with pump-induced changes of the photocurrent at pump-probe delay of $t=540$\,fs after photoexcitation at $\hbar\omega_{\text{pump}}=2$\,eV with a fluence of $F=0.4$\,mJ/cm$^2$ (right). White lines are guides to the eye highlighting the main features of the band structure. The gray-shaded area in the panel on the left-hand side indicates the momentum region where EDCs were extracted to determine the transient band position shown in Fig. \ref{figure2}a. The black arrow indicates the momentum range of the EDCs used for the determination of transient electronic temperature and chemical potential shown in Fig. \ref{figure3}. (d) Photocurrent integrated over the areas marked by the colored boxes in the right panel of Fig. \ref{figure1}c as a function of pump-probe delay together with exponential fits. The gray-shaded Gaussian in panels b and d represents the pump-probe cross-correlation (see SM \cite{SM}). The blue-shaded area in panels b and d indicates the time window where the pump-probe traces reach their extrema.}
\end{figure}

Here we use time- and angle-resolved photoemission spectroscopy (trARPES) to provide direct evidence for the occurrence of efficient CM in photoexcited Bi$_2$Se$_3$. The underlying idea is given in Figs. \ref{figure1}a and b. Figure \ref{figure1}a shows a sketch of the band structure of a typical semiconductor with the Fermi level located in the conduction band (CB) as in the case of the Bi$_2$Se$_3$ samples investigated in the present work. Photoexcitation with photon energies much bigger than the gap size triggers direct electronic transitions from the valence band (VB) into unoccupied states above the Fermi level in the CB. The excess energy $E_{exc}$ of the photoexcited electron is used to promote additional electrons from the VB into the CB. In order to reveal signatures of CM in experimental trARPES data, we propose to integrate the photocurrent over the areas indicated by the colored boxes in Fig. \ref{figure1}a to gain access to the transient population of the VB and CB, respectively. After photoexcitation, the transient population of the CB is expected to increase while the transient population of the VB is expected to decrease. In the absence of CM (dashed curves in Fig. \ref{figure1}b) the corresponding pump-probe traces will reach their respective maximum or minimum immediately after the pump pulse is gone. In the presence of CM, however, the number of electron-hole pairs will keep increasing beyond this time interval (continuous curves in Fig. \ref{figure1}b). A quantitative estimate of the CM efficiency can be obtained from the difference in amplitude between the continuous and dashed curves in Fig. \ref{figure1}b. 

Using this approach, we obtain a CM factor of $\sim2.3$ for the photoexcitation parameters used in the present study. We also analyze transient changes in binding energy that have been previously interpreted in terms of SPV effects \cite{Hajlaoui2014, Neupane2015, Sanchez2017, Ciocys2020, Ciocys2020, Michiardi2022} as well as the transient carrier distribution inside the CB to provide a complete picture of photocarrier dynamics in Bi$_2$Se$_3$. The microscopic understanding gained in the present work might find applications in novel Bi$_2$Se$_3$-based photodetectors.



Single crystals of Bi$_2$Se$_3$ were grown by melting stoichiometric mixtures of high-purity elemental Bi and Se as described in detail in \cite{Xia2009}. The samples were cleaved in ultrahigh vacuum using Kapton tape.

Our trARPES setup was based on a Titanium:Sapphire amplifier operating at a repetition rate of 1\,kHz with a pulse energy of 7\,mJ and a pulse duration of 35\,fs. 5\,mJ of output energy were used to seed an optical parametric amplifier. The signal output was frequency-doubled to provide visible pump pulses with a photon energy of $\hbar\omega_{\text{pump}}=2$\,eV. The remaining 2\,mJ of output energy were frequency-doubled and used for high harmonics generation of extreme ultraviolet (XUV) pulses in argon. A single harmonic at $\hbar\omega_{\text{probe}}=21.7$\,eV, isolated with a grating monochromator, served as a probe pulse that ejected photoelectrons from the sample. The photoelectrons passed through a hemispherical analyzer providing direct snapshots of the occupied part of the band structure of the solid under investigation.

	

In Figs. \ref{figure1}c and d we provide experimental data that we analyze as proposed above to look for the possible occurrence of CM in Bi$_2$Se$_3$. The left panel of Fig. \ref{figure1}c shows the momentum- and energy-resolved photocurrent around the $\Gamma$-point of Bi$_2$Se$_3$ measured at negative pump-probe delay before the arrival of the pump pulse. The white lines are guides to the eye that highlight the main features of the band structure: a parabolic CB, an M-shaped VB (VB$_M$), and a second V-shaped VB (VB$_V$) at lower energy. The topological surface state is not resolved, likely due to a poor photoemission cross section at $\hbar\omega_{\text{probe}}=21.7$\,eV as well as the limited signal-to-noise ratio of our trARPES data. The pump-induced changes of the photocurrent for a pump-probe delay of $t=540$\,fs after photoexcitation at $\hbar\omega_{\text{pump}}=2$\,eV with a fluence of $F=0.4$\,mJ/cm$^2$  are shown in the right panel of Fig. \ref{figure1}c. Red and blue indicate a gain and loss of photoelectrons with respect to negative pump-probe delay, respectively. In Fig. \ref{figure1}d we plot the transient change of the occupation of CB and VB$_M$, respectively, obtained by integrating the counts over the areas marked by the colored boxes in Fig. \ref{figure1}c. Continuous lines are exponential fits to the data (see Supplemental Material (SM)\cite{SM}) yielding lifetimes of $\tau=3.1\pm0.9$\,ps and $\tau=2.1\pm0.6$\,ps for electrons and holes, respectively, in good agreement with literature \cite{Sobota2012}. The gray-shaded area in Fig. \ref{figure1}c represents the pump-probe cross correlation (see SM \cite{SM}) with a full width at half maximum of FWHM\,$\sim150$\,fs.

We find that the number of electron-hole pairs reaches its maximum at a pump probe delay of $t\sim600$\,fs long after the pump pulse is gone. This is exactly what is expected to happen in the presence of CM (see sketches in Fig. \ref{figure1}a and b). By comparing the amplitudes of the pump-probe traces at $t=150$\,fs where the pump-probe cross correlation is over and at $t\sim600$\,fs where the pump-probe traces reach their extremal values, we estimate CM by a factor of $\sim1.4$ from the VB$_M$ data and $\sim3.1$ from the CB data with a mean value of $\sim2.3$. Note that the pump-probe signal in Fig. \ref{figure1}d is superimposed with coherent oscillations with a frequency of $\sim2$\,THz that we attribute to the A$_{1g}$ phonon of Bi$_2$Se$_3$ \cite{Norimatsu2013, Sobota2014b}.

\begin{figure}
\centering
\includegraphics[width=1\columnwidth]{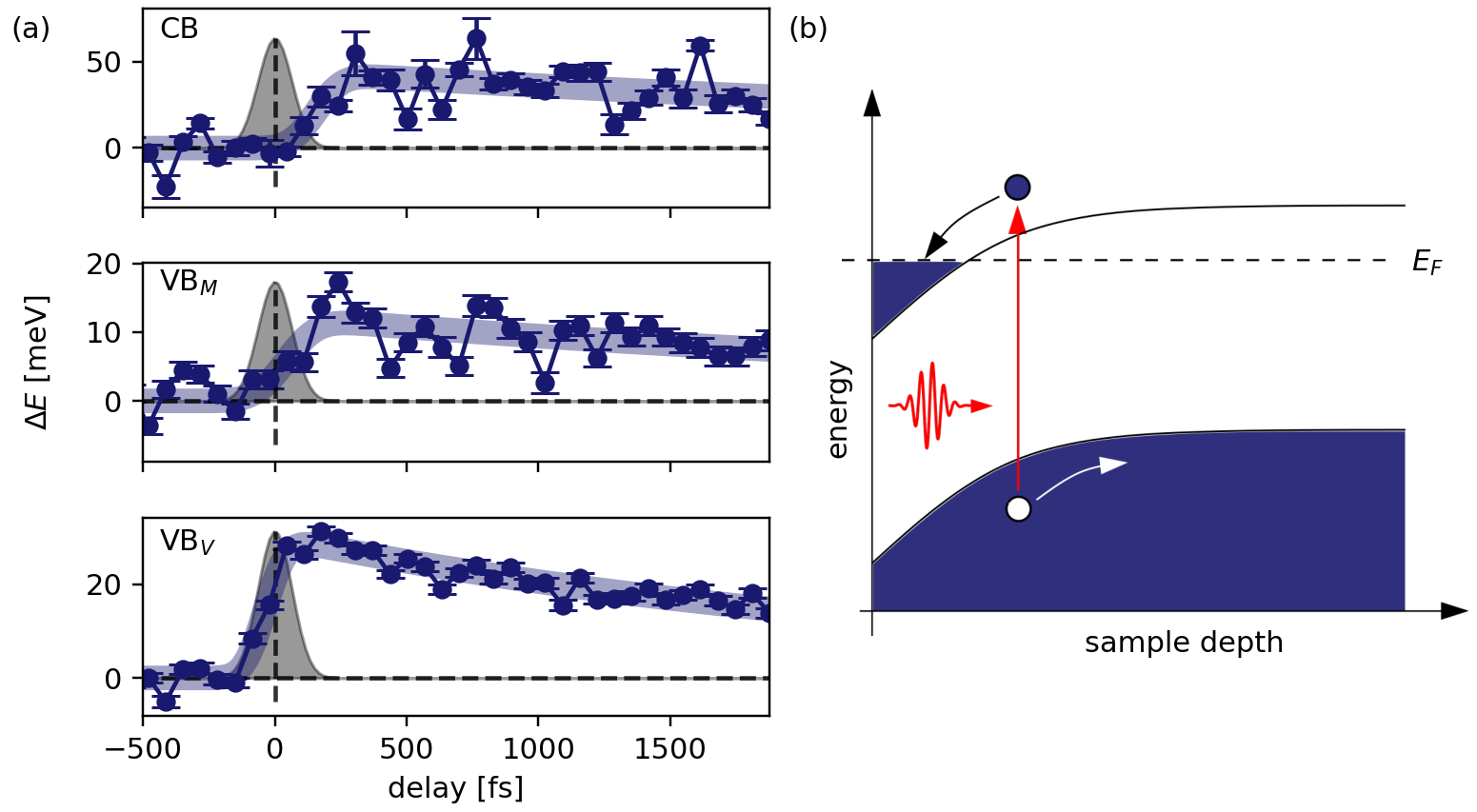}
\caption{\label{figure2} (a) Transient shifts of the CB, the M-shaped VB$_M$, and the V-shaped lower VB$_V$ as a function of pump-probe delay together with exponential fits. The gray-shaded Gaussian represents the pump-probe cross-correlation (see SM \cite{SM}). (b) Sketch explaining the origin of the SPV effect in photoexcited Bi$_2$Se$_3$.}
\end{figure}

\begin{figure}
\centering
\includegraphics[width=0.5\columnwidth]{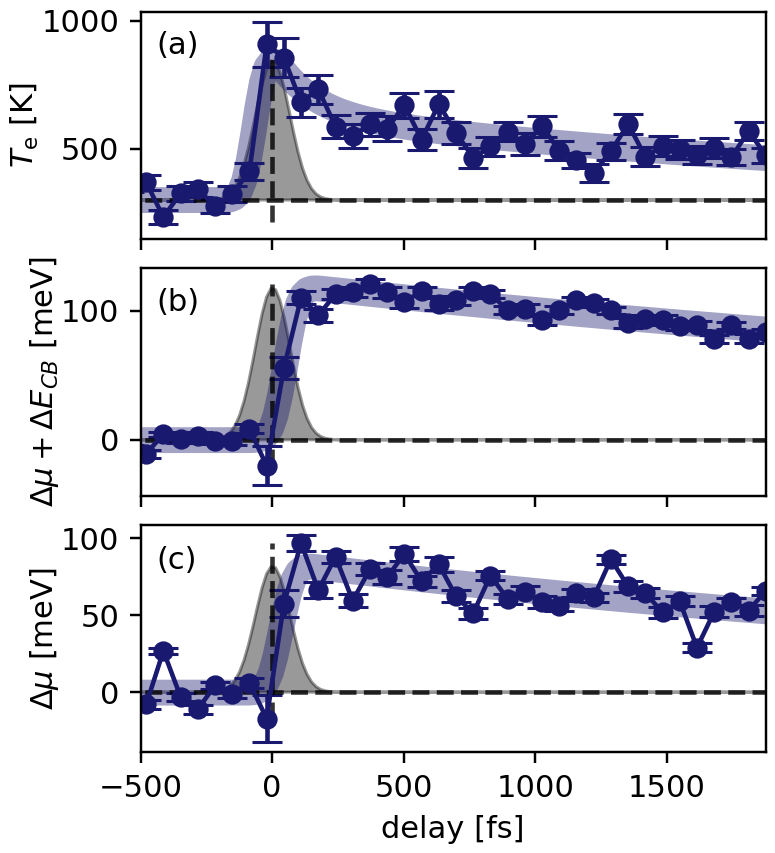}
\caption{\label{figure3} Transient carrier distribution inside the CB obtained from Fermi-Dirac fits. (a) Resulting electronic temperature, (b) position of Fermi edge, and (c) change of chemical potential as a function of pump-probe delay together with exponential fits. The gray-shaded Gaussian represents the pump-probe cross-correlation (see SM \cite{SM}).}
\end{figure}

After having established efficient CM in photodoped Bi$_2$Se$_3$ as the main result of the present work, we continue by analyzing the transient band structure (Fig. \ref{figure2}) and the transient carrier distribution in the CB (Fig. \ref{figure3}) to provide a complete picture of the non-equilibrium carrier dynamics. To gain access to the transient binding energies of CB, VB$_M$ and VB$_V$, we extract energy distribution curves (EDCs) by integrating the photocurrent over the momentum range indicated by the gray-shaded area in Fig. \ref{figure1}c. The transient peak positions are extracted from Gaussian fits for VB$_V$ and VB$_M$. In order to determine the shift of the CB, we fit the rising edge at the bottom of the CB with an error function. Further details about the fits are provided in the SM \cite{SM}. Note that these fits also allow us to extract the transient spectral weight of VB$_M$ (see SM SFig.\,3 \cite{SM}) which is found to show the same dynamics as the red pump-probe trace in Fig.\,\ref{figure1}d. The temporal evolution of the transient changes in binding energy of CB, VB$_M$ and VB$_V$ are shown in Fig. \ref{figure2}a as a function of pump-probe delay together with exponential fits (see SM \cite{SM}). All bands are found to shift up with amplitudes on the order of a few tens of meV and lifetimes of several picoseconds indicating that the surface region of the sample which is probed by trARPES becomes negatively charged. Similar shifts were observed in photodoped Bi$_2$Se$_3$ in the past and interpreted in terms of a SPV \cite{Hajlaoui2014, Neupane2015, Sanchez2017, Ciocys2020, Ciocys2020, Michiardi2022}. 

A sketch explaining the origin of this SPV effect is shown in Fig. \ref{figure2}b. In a nutshell, the presence of a metallic surface state on an insulating bulk results in the formation of a Shottky barrier and a band bending at the surface the details of which depend on the doping level \cite{Hajlaoui2014, Neupane2015, Sanchez2017, Ciocys2020, Ciocys2020}.  In our samples the bulk CB is pulled below the Fermi level in the surface region. Due to the band bending, photoexcited electrons and holes drift towards the surface and bulk of the sample, respectively, accumulating negative charges at the surface that reduce the binding energy of the states. The lifetime of these SPV effects strongly depends on the doping level of the bulk. Short-lived SPV effects with lifetimes on the order of 10\,ps as observed in the present case were previously reported for bulk conductive samples \cite{Sanchez2017, Sterzi2017, Ciocys2020}, while bulk insulating samples exhibit much longer SPV lifetimes on the order of nanoseconds to picoseconds \cite{Neupane2015, Sterzi2017, Sumida2017, Sanchez2017, Ciocys2020}.

In Figure \ref{figure3} we present our analysis of the transient carrier population of the CB. We extracted EDCs by integrating the photocurrent over the momentum range indicated by the black arrow in the left panel of Fig. \ref{figure1}c that we fitted with a Fermi-Dirac distribution. The fact that these fits work well for all pump-probe delays (see SM SFig.\,4) indicates that the electrons inside the CB thermalize on a time scale that is short compared to our temporal resolution. The resulting transient electronic temperature and the shift of the transient Fermi edge are shown in Figs. \ref{figure3}a and b, respectively, together with exponential fits (see SM \cite{SM}). The electronic temperature is found to rise to $\sim900$\,K and to cool down with a double-exponential decay with lifetimes of $\tau_1\sim150$\,fs and $\tau_2\sim3$\,ps. These values are in good agreement with literature \cite{Sobota2012, Wang2012} and are assigned to the emission of optical and acoustic phonons, respectively. The shift of the transient chemical potential (see Fig. \ref{figure3}c) defined as the energy difference between the transient Fermi edge and the CB minimum is obtained by subtracting the CB shift shown in Fig. \ref{figure2}a from the data in Fig. \ref{figure3}b. The transient chemical potential is found to increase by $\sim80$\,meV with a lifetime of $\tau\sim5$\,ps (see Fig. \ref{figure3}c). 

The shift of the chemical potential in the CB likely has three contributions: (1) a reduction due to the elevated electronic temperature, (2) an increase due to CM as observed in Fig. \ref{figure1}d, and (3) an increase due to the transient accumulation of electrons at the surface as sketched in Fig. \ref{figure2}b. The lifetime of the shift of the chemical potential is much longer than the $\tau_1$ lifetime of the electronic temperature in Fig. \ref{figure3}a but similar to the lifetime of the band populations in Fig. \ref{figure1}d and the band shifts in Fig. \ref{figure2}a. Therefore, the latter two contributions seem to dominate. Thus, the transient increase of the number of electrons in the CB is not due to CM alone. 

Therefore, to provide clear evidence for the occurrence of CM in Bi$_2$Se$_3$, we need to disentangle the contributions of CM and SVP effects to the pump-probe traces in Fig.\,\ref{figure1}d. In the absence of CM, SVP effects are expected to increase the number of photoexcited electrons in the CB and reduce the number of photoexcited holes in the VB$_M$ on a timescale determined by the drift velocity of the carriers and the thickness of the band bending region. This timescale can be extracted from the risetime of the SPV-induced band shifts in Fig.\,\ref{figure2}a that is found to be on the order of the pump-probe cross correlation. Any dynamics occurring at later times, in particular the delayed extremal values of the pump-probe traces in Fig.\,\ref{figure1}d, cannot be attributed to SPV effects. Further, SPV effects cannot account for the observed increase of the number of holes in the VB$_M$ in Fig.\,\ref{figure1}d. Thus, the occurrence of CM is essential to explain the experimentally observed dynamics. However, the CM factor of $\sim3.1$ determined from the CB data likely overestimates the true value due to additional electron accumulation at the surface from SPV effects. On the other hand, the factor of $\sim1.4$ determined from the VB$_M$ data likely underestimates the true value due to the drift of the photoexcited holes into the bulk. The average value of $\sim2.3$, however, likely presents a sensible estimate.



In summary, our trARPES results on photoexcited Bi$_2$Se$_3$ revealed an increase of the number of electron-hole pairs for timescales much longer than the pump-probe cross correlation providing direct evidence for CM. In addition, we confirmed the existence of SPV effects and the relaxation times of the hot electronic distribution inside the CB as previously discussed in literature. Starting from the present findings one can now set out to improve the CM efficiency of Bi$_2$Se$_3$ in a systematic investigation of the influence of the pump photon energy, the pump fluence, and the equilibrium doping level of the sample.  

\begin{acknowledgments}
This work received funding from the European Union’s Horizon 2020 research and innovation program under Grant Agreement No. 851280-ERC-2019-STG and from the Deutsche Forschungsgemeinschaft (DFG) via the collaborative research center CRC 1277 (project No. 314695032). J.F.K. was supported by the Arnold O. Beckman Postdoctoral Fellowship.
\end{acknowledgments}

\clearpage

\bibliography{literature}
\bibliographystyle{apsrev4-2}

\clearpage

\clearpage
\pagebreak

\section{Supplementary Information}

\subsection{estimation of pump-probe cross-correlation}
\begin{figure}[b]
\centering
\includegraphics[width=1\linewidth]{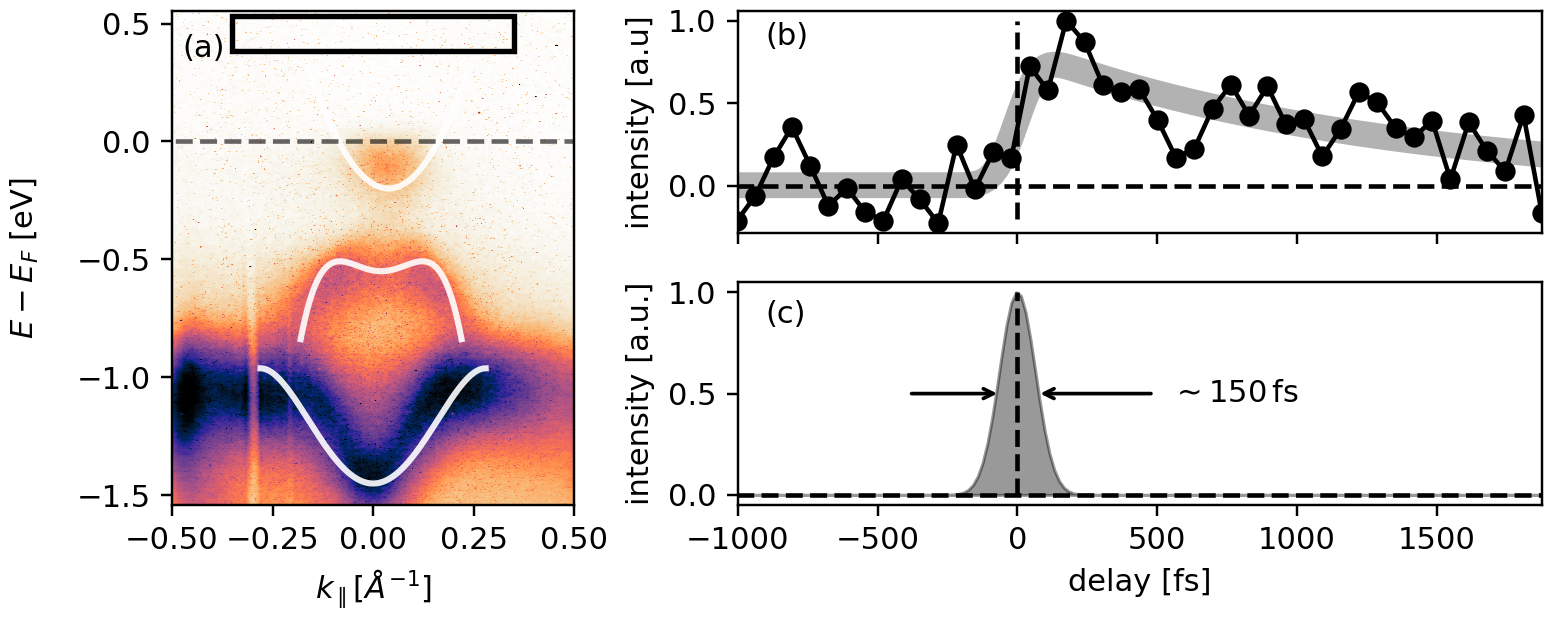}
\caption{\label{SFigure1} Pump-probe cross-correlation: (a) ARPES snapshot of Bi$_2$Se$_3$. The box indicates the area of integration for the pump-probe trace that is shown in panel (b) together with an exponential fit. (c) Pump-probe cross correlation extracted from the derivative of the rising edge of the fit in b.}
\end{figure}

The pump-probe cross-correlation used in the main text was determined as follows: First we integrated the photocurrent over the area marked by the black box in SFig.\,\ref{SFigure1}a. The resulting pump-probe trace (see SFig.\,\ref{SFigure1}b) was then fitted with an exponential decay (see below). The resulting full width at half maximum of the derivative of the rising edge of this fit provides a good estimate of the pump-probe cross-correlation, which is plotted in SFig.\,\ref{SFigure1}c. 

\subsection{determination of transient band positions}
\begin{figure}
\centering
\includegraphics[width=1\linewidth]{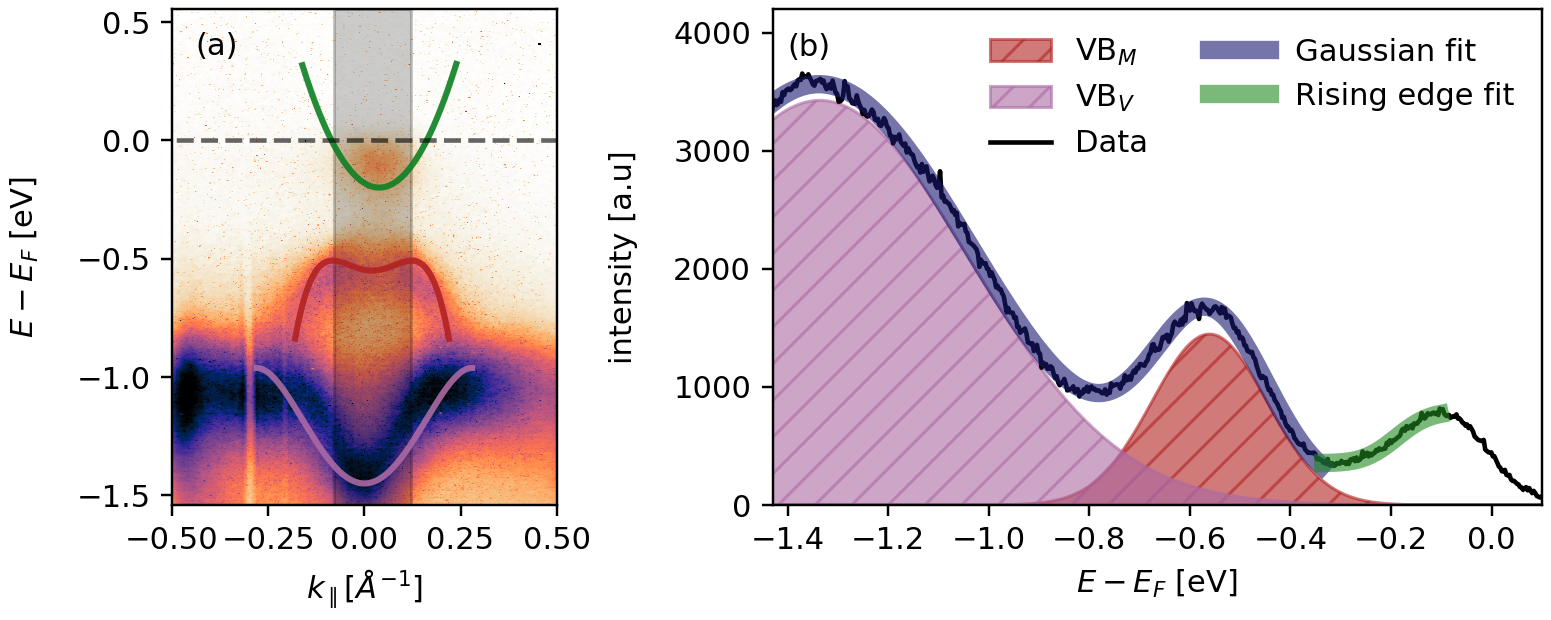}
\caption{\label{SFigure2} Determination of transient peak positions: (a) ARPES snapshot of Bi$_2$Se$_3$. The colored lines are guides to the eye that highlight the main features of the band structure: the parabolic CB in green, the M-shaped VB$_M$ in red, and the V-shaped VB$_V$ in violet. The gray-shaded area indicates the range of integration for the EDC in b. (b) EDC at negative delay together with Gaussian fit for VB$_M$ and VB$_V$ in blue and CB fit in green. The dashed areas indicate the two individual Gaussians used to fit VB$_M$ and VB$_V$.}
\end{figure}

The transient band positions were determined by extracting EDCs from the gray-shaded area in SFig.\,\ref{SFigure2}a. The positions of VB$_M$ and VB$_V$ were fitted with the sum of two Gaussians and a constant background (blue line in SFig.\,\ref{SFigure2}b). The position of the CB, on the other hand, was determined by fitting its rising edge at the band minimum with an error function (green line in SFig.\,\ref{SFigure2}b). A proper fit function for the CB, consisting of a Gaussian multiplied by the Fermi-Dirac distribution, contained too many free parameters that turned the determination of the peak position unreliable.  

\begin{figure}
	\centering
	\includegraphics[width=1\linewidth]{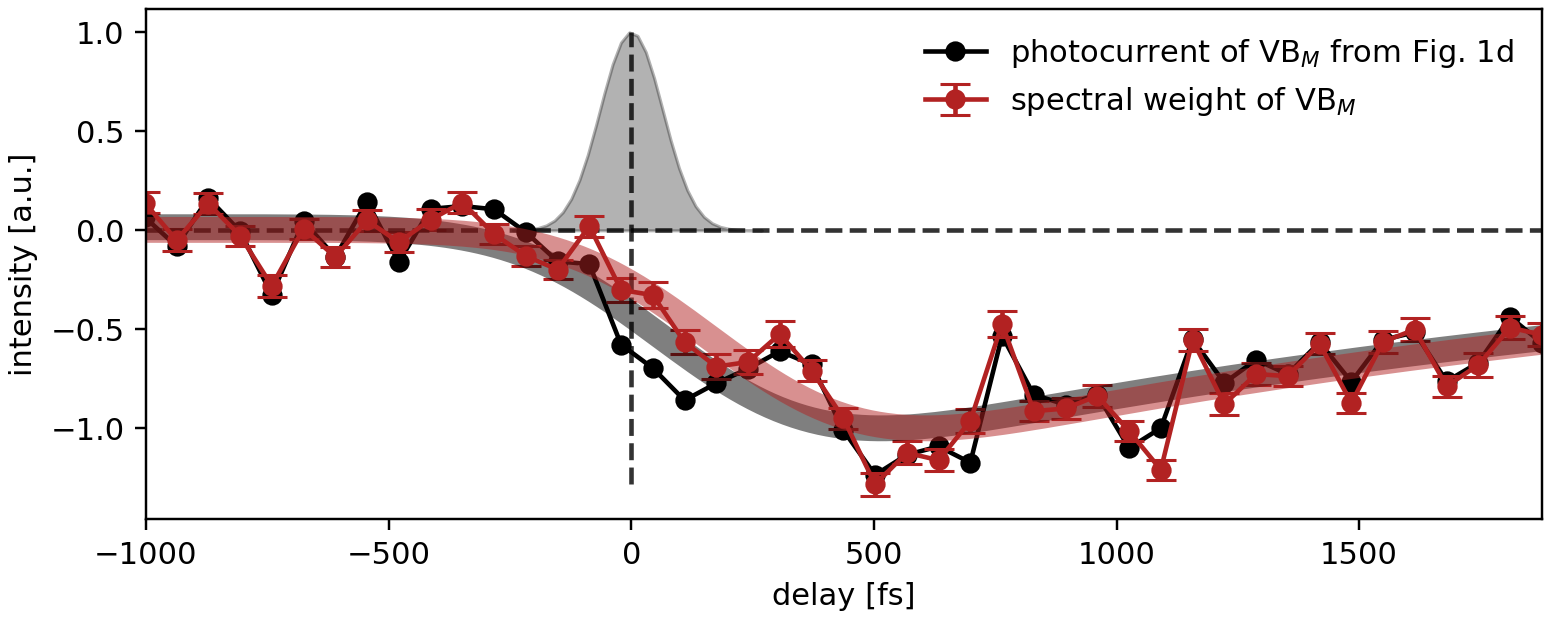}
	\caption{\label{SFigure3} Direct comparison between pump-probe trace for VB$_M$ from Fig. 1d in the main text (black) and transient spectral weight of VB$_M$ extracted from Gaussian fit (red). The gray-shaded area is the pump-probe cross correlation. Thick lines are exponential fits.}
\end{figure}

The Gaussian fits also allow us to extract the transient spectral weight of VB$_M$ (see SFig.\,\ref{SFigure3}) which is found to show the same dynamics as the red pump-probe trace in Fig.\,\ref{figure1}d of the main text.

\begin{figure}
\centering
\includegraphics[width=1\linewidth]{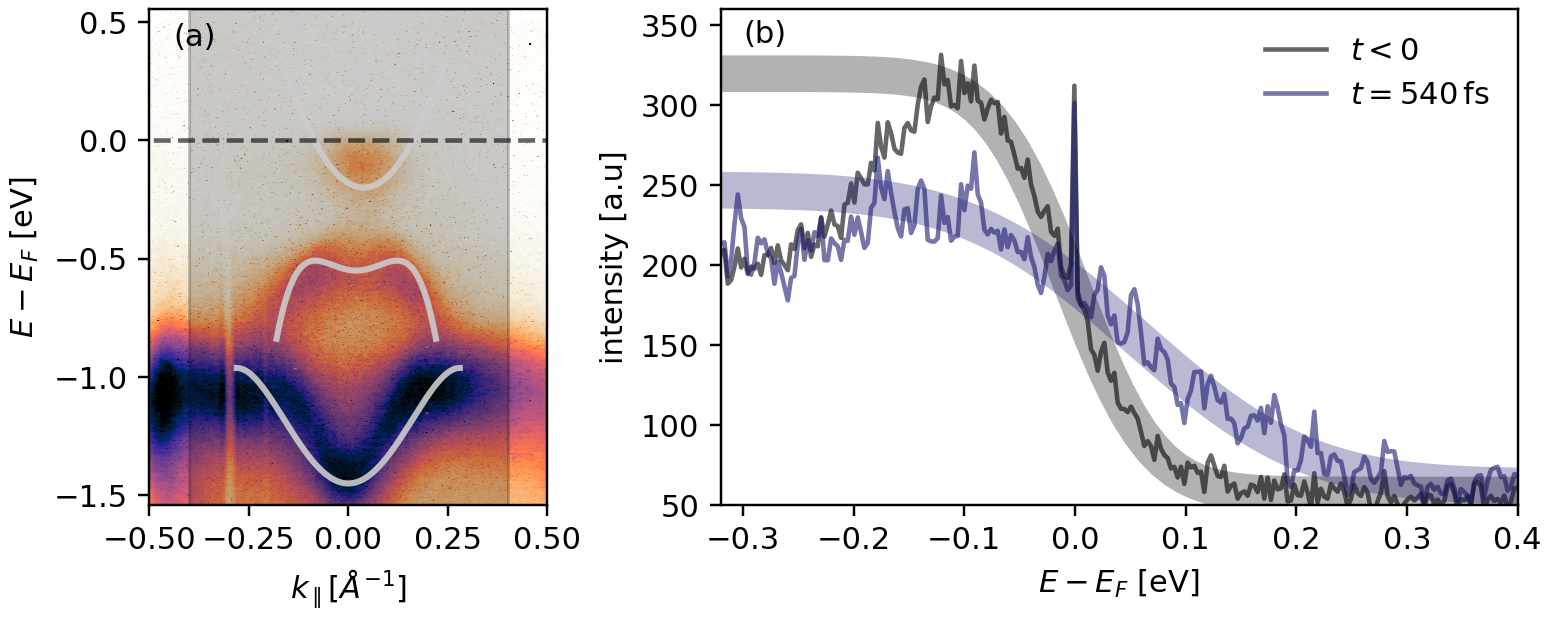}
\caption{\label{SFigure4} Fermi-Dirac fits: (a) ARPES snapshot of Bi$_2$Se$_3$. The gray-shaded area indicates the range of integration for the EDCs in b. (b) EDCs together with Fermi-Dirac fits at negative pump-probe delay and at $t=540\,$fs in gray in blue, respectively.}
\end{figure}

\subsection{exponential decay fit}
The dynamics of all transients shown in the main article were extracted using a fit function consisting of a step function $\Theta$ and a single-exponential decay, the product of which was then convolved with a Gaussian to account for the finite temporal resolution. The analytical expression is given by: 
\begin{equation}
\label{eq:RaD_fit}
f(t)=	\frac{1}{2}\bigg[ 1+\mathrm{erf}\bigg( \frac{\tau(t-t_0)-\sigma^2}{\sqrt{2}\, \sigma \tau } \bigg) \bigg]  \exp\bigg( \frac{\sigma^2 - 2\tau(t-t_0)}{2\tau^2} \bigg),
\end{equation}
where $\mathrm{erf}$ is the error function, $t_0$ is the position of the rising edge, $\sigma$ is the width of the rising edge, and $\tau$ is the time constant of the exponential decay. The full width at half maximum of the derivative of the rising edge is given by FWHM\,$=\sqrt{2\ln2}\,\sigma$.

\subsection{Fermi-Dirac fit}
To extract the electronic temperature and the position of the Fermi cutoff, we extracted EDCs from a momentum interval of $\pm0.4$\,\AA$^{-1}$ around the $\Gamma$-point (gray-shaded area in SFig.\,\ref{SFigure4}a). The resulting EDCs were fitted by a Fermi-Dirac distribution convolved with a Gaussian to account for the finite energy resolution (see SFig.\,\ref{SFigure4}b).

\end{document}